\documentclass[12pt]{iopart}
\usepackage{latexsym}

\def\tr{{\rm tr}}
\def\ket#1{\mid~\!\!\!{#1}~\!\!\rangle}
\def\bra#1{\langle~\!\!{#1}~\!\!\!\mid}

\begin{document}\jl{1}

\title[\bf Information in Multipartite States
and Strong Subadditivity of Entropy] {\bf
On Mutual Information\\ in Multipartite
Quantum States and\\ Equality in Strong
Subadditivity of Entropy}
\author{F Herbut\footnote[1]{E-mail:
fedorh@infosky.net}}
\address{Serbian Academy of
Sciences and Arts, Knez Mihajlova 35,
11000 Belgrade, Serbia}

\date{\today}

\begin{abstract}
The challenge of equality in the strong
subadditivity inequality of entropy is
approached via a general additivity of
correlation information in terms of
nonoverlapping clusters of subsystems in
multipartite states (density operators).
A family of tripartite states satisfying
equality is derived.
\end{abstract}

\maketitle

\normalsize

\rm Two, at first glance unrelated,
concepts are investigated in this
article: the correlation information
connected with nonoverlapping composite
subsystems (clusters) of a multipartite
quantum system and equality in strong
subadditivity (SSA) of entropy. It is
shown that the first notion is useful for
treating the second one. Needless to
state that both concepts are important
for quantum information theory.

Let $\rho_{1\dots N}$ be a multipartite
state (density matrix), $S_{1\dots
N}\equiv S(\rho_{1\dots N})\equiv -\tr
[\rho_{1\dots N}log(\rho_{1\dots N})]$
the corresponding quantum entropy,
$\rho_1\equiv \tr_{2\dots N}(\rho_{1\dots
N})$, $\rho_2$ etc. the reductions, and
$S_1$, $S_2$ etc. the corresponding
entropies.

The well known subadditivity of entropy
claims that always $S_{12}\leq S_1+S_2$,
and that one has equality if and only if
$\rho_{ 12}=\rho_1\otimes \rho_2$
\cite{Wehrl}. This generalizes to $N$
subsystems.

{\bf Lemma 1.} {\it For all states
$\rho_{1\dots N}$ and for $N\geq 2$  the
subadditivity
$$S_{1\dots N}\leq \sum_{n=1}^NS_n
\eqno{(1)}$$ is valid, and one has
equality if and only if $\rho_{1\dots
N}=(\Pi^{\otimes })_{n=1}^N \rho_n$, i.
e., if all subsystems are uncorrelated.}

{\bf Proof.} If the lemma is valid for
$(N-1)$ subsystems, i. e., if $S_{1\dots
(N-1)}\leq \sum_{n=1}^{(N-1)}S_n$, and
one has equality if and only if
$\rho_{1\dots (N-1)}=(\Pi^{\otimes
})_{n=1 }^{(N-1)}\rho_n$, then it is
valid also for $N$ subsystems. This is so
because the first $(N-1)$ subsystems can
be understood as one (composite)
subsystem. Then subadditivity for two
subsystems implies $S_{1\dots N}\leq
S_{1\dots (N-1)}+S_N$. This inequality in
conjunction with the preceding one leads
to $S_{1\dots N}\leq \sum_{n=1}^NS_n$.
One has equality if and only if both
inequalities preceding the last one are
equalities. These are equivalent to
$\rho_{1\dots N}=[(\Pi^{\otimes
})_{n=1}^{(N-1)}\rho_n] \otimes \rho_N$.
Since the claim is valid for $N=2$, by
total induction it is valid for all
$N\geq 2$.\hfill $\Box$

The nonnegative quantity
$$I_{1\dots N}\equiv
\sum_{n=1}^NS_n-S_{1\dots N}\eqno{(2)}$$
is called the {\it correlation
information} (contained) in $\rho_{1\dots
N}$. For $N=2$, it is called (quantum)
{\it mutual information}. The correlation
information in an $N$-partite state is
positive if $\rho_{1\dots N}$ is in any
way different from the tensor product of
all subsystem states.

Let
$$\Pi:\qquad \{1\dots
N\}=\sum_{k=1}^KC_k\eqno{(3)}$$ be an
arbitrary partitioning of the set
$\{1\dots N\}$ into classes, physically,
clusters, each consisting of some of the
subsystems $1,\enskip 2,\enskip \dots
,\enskip N$. (Note that the clusters are
nonoverlapping in the subsystems.) Let
$\rho_{C_k}$ be the reduced density
matrix corresponding to the $k-th$
cluster, obtained by tracing out in
$\rho_{1\dots N}$ all subsystems except
those belonging to the class $C_k$. Let,
further, $S_{C_k}$ be the entropy of this
density matrix. Lemma 1 in application to
the clusters implies
$$S_{C_k} \leq \sum_{n\in
C_k}S_n\eqno{(4a)}$$ with equality if and
only if all subsystems in the cluster are
uncorrelated.

Let the correlation information in the
cluster $C_k$ be
$$I_{C_k}\equiv \sum_{n\in
C_k}S_n-S_{C_k}.\eqno{(4b)}$$ We call it
the {\it within-the-cluster correlation
information.} (Note that if $C_k=\{n\}$,
then $I_{C_k}=S_n-S_n=0$.) The
correlation information in a composite
cluster is zero if and only if all
subsystems in the cluster are
uncorrelated.

Further, since the clusters can be
understood as (composite) subsystems, (1)
implies
$$S_{1\dots N}\leq \sum_{k=1}^KS_{C_k},
\eqno{(5a)}$$ and one has equality if and
only if all clusters are uncorrelated
with each other.

Finally, let
$$I_{\Pi}\equiv \sum_{k=1}^KS_{C_k}
-S_{1\dots N}\eqno{(5b)}$$ be the {\it
among-the-clusters correlation
information}. It is positive if and only
if there is any correlation among the
clusters. If $K=N$, i. e., if all
clusters in the partitioning (3) contain
only one subsystem, then (5b) has the
special form (2).

Now, we formulate the theorem on {\it
cluster additivity of correlation
information}.

{\bf Theorem 1.} {\it For every N-partite
state $\rho_{1\dots N}$ the following
additivity is valid:
$$I_{1\dots
N}=I_{\Pi}+\sum_{k=1}^KI_{C_k}.\eqno{(6)}$$
In words, the total correlation
information is the sum of the
among-the-cluster one, and the sum of the
within-the-clusters ones summed over all
clusters.}

Note that (6) is valid for every
partitioning $\Pi$.

{\bf Proof.} Adding and subtracting
$\sum_{k=1}^KS_{C_k}$ on the RHS of (2),
one obtains $$I_{1\dots
N}=\Big(\sum_{k=1}^KS_{C_k}-S_{1\dots N}
\Big)+\Big(\sum_{k=1}^K(\sum_{n\in
C_k}S_n-S_{C_k})\Big),$$ which, on
account of (5b) and (4b), gives
(6).\hfill $\Box$

The theorem is a rare statement of great
generality that is harder to state than
to prove. It implies a useful {\it
corollary on successive binary steps}.

{\bf Corollary 1.} {\it One can take
$K=2$, then within each cluster repeat
this procedure etc. In this way
$I_{1\dots N}$ is evaluated in terms of
binary steps; each step giving a term
that is a quantum mutual information.}

For $N=3$ both the theorem and the
corollary enable one only to make a
one-step binary partition; but this can
be done in three ways: $\Pi_1:\enskip
\{123\}=\{1\}+\{23\}$, $\Pi_2:\enskip
\{123\} =\{2\}+\{13\}$, and $\Pi_3:
\enskip \{123\}=\{3\}+\{12\}$. We will
write $I_{\Pi_1}$ as $I_{1,23}$ etc. to
display the fact that one is dealing with
the mutual information between subsystem
$1$ and the cluster $\{23\}$ etc.

Let us turn to the {\it strong
subadditivity} (SSA) of entropy for
tripartite systems. Its intuitively most
appealing form is $$I_{12}\leq I_{1,23}
\eqno{(7)}$$ etc. (See \cite{N-Ch},
Theorem 11.15(2), pp. 522-523. Nielsen
and Chuang state only necessity of (7),
but prove its equivalence with the
standard form of SSA.)

For $N$-partite systems, SSA has {\it the
general form} which says: Mutual
information between two nonoverlapping
clusters never increases discarding any
number of subsystems from any of the
clusters. (It is straightforward to
establish equivalence between the general
form and (7).)

Strong subadditivity is an inequality. It
is interesting to see when it is an
equality. There exist sufficient and
necessary conditions for SSA equality in
the literature, cf \cite{Ruskai},
\cite{Petz1}, \cite{Petz2} (but they are
not easily workable).

The simplest concrete example
(\cite{Ruskai}, pp. 4361-4362) is the
following:
$$\rho_{123}\equiv \rho_{12}\otimes
\rho_3.\eqno{(8)}$$ The cluster
additivity then gives utilizing $\Pi_3$:
$I_{123}=I_{12,3}+I_{12}=0+I_{12}$. On
the other hand, the partition $\Pi_1$
gives
$$I_{123}=I_{1,23}+I_{23}=I_{1,23}+0.
\eqno{(9a)}$$ Altogether,
$$I_{12}=I_{1,23},\eqno{(9b)}$$ i. e., we have an
equality in SSA (cf (7)). One can
generalize this.

{\bf Corollary 2.} {\it If $\rho_{1\dots
N}=\rho_{1\dots M}\otimes
\rho_{(M+1)\dots N}$, then $I_{C_k,C_l}=
I_{C_k,C_l'}$, where $C_l$ is a cluster
containing all subsystems $(M+1)\dots N$
and at least one subsystem besides them,
$C_k$ is a cluster nonoverlapping with
$C_l$, and $C_l'$ is obtained from $C_l$
by discarding any number of the
subsystems $(M+1)\dots N$.}

{\bf Proof.} Let $\bar C_l$ be the
cluster obtained from $C_l$ be removing
all subsystems $(M+1)\dots N$. Then, on
account of the fact that clusters can be
viewed as (composite) subsystems, (9b)
implies $$I_{C_k,\bar C_l}=I_{C_k,C_l}.
$$ On the other hand, the general form of
inequality (7)
leads to $$I_{C_k,\bar C_l}\leq
I_{C_k,C_l'},$$ and also to
$$I_{C_k,C_l'}\leq I_{C_k,C_l}.
$$ The two inequalities and the
preceding equality finally bear out the
claim.\hfill $\Box$

In \cite{Ruskai} (p. 4362) it was stated
that no other special case has been
found. The derivation that follows is a
reaction to this challenge.

Since in case (8) $I_{23}=0$, and the
first equality in (9a) is generally
valid, one might think that this lack of
correlations between subsystems $2$ and
$3$ is the crucial point. This would be a
wrong conjecture. We derive now a family
of cases of SSA equality (9b) in which
$I_{23}>0$.

First we define the notion of  a mixture
(or a state decomposition) that is {\it
biorthogonal}.

{\bf Definition 1.} {\it A state
decomposition
$$\rho_{12}=\sum_kw_k\rho_{12}^k\eqno{(10a)}$$
$(\forall k:\enskip w_k>0,\enskip \rho_{1
2}^k>0,\enskip \tr \rho_{ 12}^k=1;\enskip
\sum_kw_k=1)$ is biorthogonal if, in
terms of the reductions of $\rho_{
12}^k$},
$$\forall k\not= k':\qquad
\rho_s^k\rho_s^{k'}=0,\quad
s=1,2.\eqno{(10b)}$$

Now we define a family of states
$\rho_{123}$ {\it satisfying the SSA
equality} (9b).

{\bf Theorem 2.} {\it Let $$\rho_{123}=
\sum_kw_k\rho_{123}^k\eqno{(11)}$$ be a
mixture of states such that, tracing out
subsystem $3$, one obtains a biorthogonal
state decomposition. Then, if SSA
equality (9b) is valid for each state
$\rho_{123}^k$ in the mixture (11), then
it is valid also for $\rho_{123}$.
Further, if there are at least two terms
in the decomposition, then $I_{23}>0$ for
$\rho_{123}$ unless (11) is a special
case of (8).}

To prove the theorem, we need four
auxiliary lemmas. The first is concerned
with {\it implied biorthogonality} in
(11).

{\bf Lemma 2.} {\it If one views the
tripartite system as a bipartite one, in
particular as $\{123\}=\{1\}+\{23\}$,
decomposition (11) is biorthogonal.}

{\bf Proof.} Let $\forall k:\enskip
\rho_2^k\equiv \tr_{13}\rho^k_{123}
=\sum_ir_i^k\ket{ki}_2\bra{ki}_2$ be
spectral decompositions in terms of
positive eigenvalues. Substitution in
(10b) for $s=2$ gives
$$\forall k\not= k':\quad
\sum_i\sum_{i'}r^k_ir^{k'}_{i'}\ket{ki}_2
\bra{ki}_2\ket{k'i'}_2\bra{k'i'}_2=0.$$
This implies $$\forall k\not= k',\enskip
\forall i,\enskip \forall i':\quad
\bra{ki}_2\ket{k'i'}_2=0.$$ If
$R_2^k\equiv \sum_i\ket{ki}_2\bra{ki}_2$
are the range projectors of $\rho_2^k$,
then one further has $$\forall k\not=
k':\qquad R_2^kR_2^{k'}=0. \eqno{(12)}$$

One can always write $\rho_{23}^k=R_2^k
\rho_{23}^k=\rho_{23}^kR_2^k$ (cf
relation that is below (12a) in
\cite{FHMD}). Hence, $$\forall k\not=
k':\quad \tr
[\rho_{23}^k\rho_{23}^{k'}]=\tr
[R_2^k\rho_{23}^k\rho_{23}^{k'}R_2^{k'}]
=\tr
[(R_2^{k'}R_2^k)\rho_{23}^k\rho_{23}^{k'}]
=0.$$ (Relation (12) has been utilized.)

One can further write $$0=\tr
[\rho_{23}^k\rho_{23}^{k'}]= \tr
[(\rho_{23}^k)^{1/2}\rho_{23}^{k'}
(\rho_{23}^k)^{1/2}].$$ This implies
$[(\rho_{23}^k)^{1/2}\rho_{23}^{k'}
(\rho_{23}^k)^{1/2}]=0$ because the
operator is positive (cf Lemma A.1. in
\cite{FHFP96}). Further, $\rho_{23}^{k'}
(\rho_{23}^k)^{1/2}=0$ (cf Lemma A.2. in
\cite{FHFP96}). Multiplying this from the
right by $(\rho_{23}^k)^{1/2}$, one
finally obtains
$$\forall k\not= k':\quad \rho_{23}^k
\rho_{23}^{k'}=0.\eqno{(13)}$$ This, in
conjunction with with (10b) for $s=1$,
completes the proof.\hfill $\Box$

Thew next lemma concerns {\it the mixing
property of mutual information for
biorthogonal mixtures}.

{\bf Lemma 3.} {\it Mutual information of
a biorthogonal mixture with weights $w_k$
equals the sum of the Shannon entropy
$H(w_k)\equiv -\sum_kw_klogw_k$ and the
average mutual informations of the states
that are mixed.}

This lemma was proved in \cite{Roleof}
(see Lemma 9 there).

Next we state and prove {\it the
generalized mixing property of entropy of
any mixture}, which is known, but perhaps
not well known.

{\bf Lemma 4.} {\it Let $$\rho
=\sum_kw_k\rho^k\eqno{(14a)}$$ be any
state decomposition. Then the following
entropy decomposition is valid: $$S(\rho
)= \sum_k [w_kS(\rho^k||\rho
)]+\sum_k[w_kS(\rho^k)],\eqno{(14b)}$$
where $S(\rho ||\sigma )\equiv \tr (\rho
log\rho )-\tr (\rho log\sigma )]$ is the
relative entropy of the corresponding
states (if the support of $\sigma$
contains that of $\rho$).}

{\bf Proof.} First we must prove that
$$\forall k:\qquad supp(\rho^k)\subseteq
supp(\rho ).\eqno{(15)}$$ (By "support"
one means the subspace that is the
topological closure of the range.) In
view of the fact that the support of a
density matrix $\sigma$ is spanned by any
set of pure states into which $\sigma$
can be decomposed (cf Appendix(ii) in
\cite{MixingProp}), one should decompose
each $\rho^k$ into pure states and
substitute in (14a). Then (15) obviously
follows.

By substituting (14a) in part of (14b)
(though not everywhere), one obtains:
$$S(\rho )-\sum_k[w_kS(\rho^k)]=-\tr
\{\sum_k[w_k \rho^klog(\rho
)]\}+\sum_k\{w_k\tr [\rho^k
log(\rho^k)]\}=$$ $$\sum_k\{w_k[-\tr
(\rho^klog\rho )+\tr
(\rho^klog\rho^k)]\}=\sum_k[w_k
S(\rho^k||\rho )].$$\hfill $\Box$

{\bf Remark 1.} {\it In the special case
when (14a) is an orthogonal state
decomposition, i.e., when $\forall k\not=
k':\enskip \rho^k\rho^{k'}=0$, then (14b)
takes on the well known special form
$$S(\rho )=
H(w_k)+\sum_k[w_kS(\rho^k)],\eqno{(16)}$$
where $H(w_k)$ is the Shannon entropy of
the probability distribution
$\{w_k:\forall k\}$. One refers to (16)
as the mixing property of entropy}
(\cite{Wehrl}).

Next we define the concept of {\it a
monoorthogonal mixture}.

{\bf Definition 2.} {\it A state
decomposition
$$\rho_{23}=\sum_kw_k\rho_{23}^k,\eqno{(17a)}$$
is called monoorthogonal in subsystem $2$
if one has $$\forall k\not= k':\qquad
\rho_2^k\rho_2^{k'}=0\eqno{(17b)}$$ in
terms of the corresponding reductions.}

Now we state a lemma on {\it the mixing
property of mutual information for
monoorthogonal mixtures.}

{\bf Lemma 5.} {\it If one has a mixture
(17a) monoorthogonal in subsystem $2$,
the mutual information $I_{23}$ of the
decomposed state can be written as
follows:}
$$I_{23}=\sum_kw_kS(\rho_3^k||\rho_3)+
\sum_kw_kI_{23}^k.\eqno{(18)}$$

{\bf Proof.} The argument is
straightforward in view of (17a) and
(17b) with the help of (16):
$$I_{23}\equiv S_2+S_3 -S_{23}=
\{H(w_k)+\sum_k[w_kS(\rho_2^k)]\} +$$
$$\{\sum_k[w_kS(\rho^k_3||\rho_3)]
+\sum_k[w_kS(\rho^k_3)]\} -\{H(w_k)+
\sum_k[w_kS(\rho^k_{23})]\}.$$ After
cancellation and substitution of
$I_{23}^k$, the claimed relation (18)
ensues.\hfill $\Box$

{\bf Proof} of the theorem: On account of
the implied biorthogonality (Lemma 2), we
can apply the mixing property of Lemma 3
to decomposition (11). Hence
$$I_{1,23}=H(w_k)+ \sum_kw_kI_{1,23}^k.$$
Lemma 3 can also be applied to
decomposition (10a) when it is obtained
by tracing out subsystem $3$ in (11):
$$I_{12}= H(w_k)+\sum_kw_kI_{12}^k.$$
Since by assumption $\forall k:\enskip
I_{12}^k= I_{1,23}^k$, the last two
relations bear out the first claim of the
theorem.

Finally, a glance at (18) reveals that
for $K\geq 2$, one can have $I_{23}=0$ if
and only if $\forall k:\enskip
\rho_{123}^k=\rho_{12}^k\otimes \rho_3$.
This would reduce it to case (8).\hfill
$\Box$

{\bf Remark 2.} {\it Utilizing in Theorem
2 (8) for each state $\rho_{123}^k$, but
possibly with distinct factor states for
different values of $k$, one obtains
various concrete states $\rho_{123}$
satisfying (9b).}

Let us return to inequality (7). There
are 6 distinct inequalities of this type
(obtained from (7) by permutations). Each
of them defines a nonnegative {\it excess
in mutual information}, e. g.,
$(I_{1,23}-I_{12})$. Two by two of the 6
excesses are equal. For example
$$I_{1,23}-I_{12}=I_{12,3}-I_{23}. \eqno{(19)}$$
Equality (19) is obvious if rewritten as
$$I_{1,23}+I_{23}=I_{12,3}+I_{12},$$ when
it is an instance of the cluster
additivity of correlation information in
the tripartite system. Zero excess is the
same thing as equality in SSA of entropy.
Thus, (9b) gives rise to
$$I_{23}=I_{12,3}\enskip ,\eqno{(20)}$$ and {\it
vice versa}.

When (9b) is valid, relation (20) is a
new equality in SSA. In case (8), (20) is
trivial, because subsystem $3$ has zero
mutual information both with subsystems
$2$ and $12$, and then also the excess is
zero. In the case treated in Theorem 2,
(20) is a, perhaps even surprising, {\it
new result}.

Finally, let us see if in the case
defined in Theorem 2 one can, in the
spirit of Corollary 2, replace subsystem
$3$  by a composite system, a cluster,
and discard not the whole cluster, but
only part of it, and still have no
decrease in mutual information with
another (nonoverlapping) cluster. An
affirmative answer follows from realizing
that in the proof of Corollary 2 all that
was used was the possibility of
discarding the whole cluster (equality
$I_{C_k,\bar C_l}= I_{C_k,C_l}$), and two
SSA inequalities. All three are valid
also in the present case.

In conclusion, one may say that the
general cluster additivity of correlation
information (Theorem 1), used through
successive binary partitionings
(Corollary 1), gave a useful view of
correlations. It made possible generating
new equalities (cf (20)), as well as
generalization to clusters (cf Remark 2
and the preceding passage). The main
result is the family of states
$\rho_{123}$ satisfying equality in SSA
of entropy (Theorem 2).\\

\end{document}